\documentclass[fleqn,usenatbib,useAMS]{mnras}

\usepackage{newtxtext,newtxmath}

\usepackage[T1]{fontenc}

\DeclareRobustCommand{\VAN}[3]{#2}
\let\VANthebibliography\thebibliography
\def\thebibliography{\DeclareRobustCommand{\VAN}[3]{##3}\VANthebibliography}

\usepackage{graphicx}	
\usepackage{subfigure}
\usepackage{amsmath}	
\usepackage{multirow}

\usepackage{amssymb}	
\usepackage{color}

 \title[]{ Redshift Evolution and Non-Universal Dispersion of Quasar Luminosity Correlation}

\author[Zhuoyang Li et al.]{
Zhuoyang Li,
Lu Huang\thanks{E-mail: huanglu37@mail2.sysu.edu.cn},
Junchao Wang
\\
$^{1}$School of Physics and Astronomy, Sun Yat-Sen University, 2 Daxue Road, Tangjia, Zhuhai, 519082, P.R.China\\
}

\date{Accepted 2022 September 22. Received 2022 August 22; in original form 2022 May 24}

\pubyear{2015}

\begin{document}
\label{firstpage}
\pagerange{\pageref{firstpage}--\pageref{lastpage}}
\maketitle


\begin{abstract}
The standard $\Lambda$CDM model is recently reported to deviate from the high-redshift Hubble diagram of type Ia supernovae (SNe) and quasars (QSOs) at $\sim4\sigma$ confidence level. In this work, we combine the PAge approximation (a nearly model-independent parameterization) and a high-quality QSO sample to search for the origins of the deviation. By visualizing the $\Lambda$CDM model and the marginalized $3\sigma$ constraints of  SNe+QSOs into PAge space, we confirm that the SNe+QSOs constraints in both flat and non-flat PAge cases are in remarkable tension with the standard $\Lambda$CDM cosmology. Next, we investigate the tension from the perspective of redshift-evolution effects. We find that the QSO correlation coefficient $\gamma$ calibrated by SNe+low-z QSOs and SNe+high-z QSOs shows $\sim2.7\sigma$ and $\sim4\sigma$ tensions in flat and non-flat universes, respectively. The tensions for intrinsic dispersion $\delta$ between different data sets are found to be $>4\sigma$ in both flat and non-flat cases. These results indicate that the QSO luminosity correlation suffers from significant redshift evolution and non-universal intrinsic dispersion. Using a redshift-dependence correlation to build QSO Hubble diagram could lead to biases. Thus, the $\sim4\sigma$ deviation from the standard $\Lambda$CDM probably originates from the redshift-evolution effects and non-universal dispersion of the QSO luminosity correlation rather than new physics.

\end{abstract}




\section{Introduction}

The phenomenon of cosmic accelerating expansion is firstly indicated by observing the extra dimming of the high-redshift Type Ia supernovae (SNe)~\citep{SupernovaCosmologyProject:1997czu, SupernovaSearchTeam:1998fmf, SupernovaSearchTeam:1998bnz, SupernovaCosmologyProject:1998vns}. A widely accepted explanation of this mysterious phenomenon is that a hypothetical dark energy component with negative pressure drives the homogeneous and isotropic universe to accelerate. Interpreting dark energy as a cosmological constant $\Lambda$ and assuming the validity of general relativity at all scales and epochs, the standard $\Lambda$ cold dark matter ($\Lambda$CDM) model has achieved remarkable success in agreeing with a great majority of cosmological observational measurements~\citep{Pan-STARRS1:2017jku, eBOSS:2020yzd, Jimenez:2001gg, DES:2021wwk, Planck:2018vyg}. However, these modern cosmological measurements are restricted to either the low redshift range ($0\le z\le2.33$) or the high redshift range ($z\sim1100$). The cosmic expansion history is still poorly explored in the redshift interval ($2.33<z<1100$) which is very essential for studying the dark energy models beyond the typical $\Lambda$CDM physic.

As the most luminous and persistent energy sources in our Universe, quasars (QSOs) serve as a potential candidate for high-redshift cosmological tests which can be detected up to redshift $\sim7.64$~\citep{Yang:2021imt, Wang:2021}. Several empirical correlations between spectral features and luminosity have been proposed to enable QSOs as competitive cosmological tools ~\citep{1977ApJ...214..679B, Watson:2011um, LaFranca:2014eba, SEAMBH:2014nmr}. Particularly, the most investigated and best constructed QSO luminosity correlation is the observed non-linear correlation between  ultraviolet ($L_{\rm UV}$ at 2500\AA) and X-ray ($L_X$ at 2 $\rm keV$) luminosity which is firstly proposed in ~\citet{1979ApJ...234L...9T,10.1007/978-94-010-9949-3_8,1986ApJ...305...57T} and subsequently developed in~\citep{Risaliti:2015zla,2016ApJ...819..154L,2017A&A...602A..79L,Risaliti:2018reu,Lusso:2018qor,Salvestrini:2019thn,Lusso:2020pdb,Bisogni:2021hue}. Although the detailed physical mechanism of the $L_{\rm UV}-L_X$ correlation still remains unknown~\citep{1991ApJ...380L..51H,1993ApJ...413..507H,Ghisellini:1994wx,2000ApJ...530L..65N,Merloni:2003dc,2019A&A...628A.135A}, the authors in~\citet{Lusso:2020pdb} have minimized all of the possibles systematic effects and proven the stability of this QSO luminosity correlation. Based on the $L_{\rm UV}-L_X$ correlation, \citet{Risaliti:2018reu} utilized a new technique to model-independently build the QSO Hubble diagram and extend it to $z\sim5.5$. A good agreement is found between the constructed QSO Hubble diagram and the $\Lambda$CDM model at $z<1.4$ while a $\sim 4\sigma$ deviation emerges at higher redshift range. Two follow-up works further confirm this significant deviation with more precise approaches and cleaner QSO samples~\citep{Lusso:2019akb,Lusso:2020pdb}. Since then,  the deviation between the high-redshift QSOs Hubble diagram and the $\Lambda$CDM model arises heated debates~\citep{Melia:2019nev,Yang:2019vgk,Velten:2019vwo,Mehrabi:2020zau,Zheng:2021oeq,Lian:2021tca,Colgain:2022nlb,Li:2021onq,Khadka:2020tlm,Khadka:2021xcc}.

\citet{Velten:2019vwo} used a model-independent estimator to test the robustness of the QSO Hubble diagram. Their result suggests that the QSO data can not be used as a reliable cosmological tool because it even fails to state the cosmic accelerating expansion phase. \citet{Yang:2019vgk} claimed that the model-independent approach developed in~\citet{Risaliti:2018reu} failed to recover the high-z cosmic expansion history of the flat $\Lambda$CDM model, which undermined the $\sim 4\sigma$ deviation. Using the Gaussian process and a combination of SNIa, Quasars and gamma-ray burst data, \citet{Mehrabi:2020zau} found a less significant tension. They argued that the amount of the deviation might be affected by the choice of the kernel function. All these works challenge the claimed $\sim 4\sigma$ deviation. However, the main cause of the deviation is still not found.

Our aim in this present work is to search for the possible origins of the significant deviation between the $\Lambda$CDM model and the high-z Hubble diagram of  SNe+QSOs. The key is to adopt model-independent approaches and independent samples. To avoid model dependence, we perform our analyses in PAge approximation (Parameterization based on cosmic Age) which is a general approximation of many late-time cosmological models and a nearly model-independent framework~\citep{Huang:2020mub,Luo:2020ufj,Huang:2020evj,Huang:2021aku,Huang:2021tvo,Cai:2021weh,Cai:2022dkh, Huang:2022txw}. In addition, we take the most up-to-date QSO samples compiled by~\citet{Lusso:2020pdb} as our data set.

Our work is organized as follows. We present the advantages of the PAge approximation in the next Section~\ref{Sec:cosmology}. The data and methodology are briefly introduced in Section~\ref{Sec:data-method}. The detailed results are shown in Section~\ref{Sec:results}. In the last Section~\ref{Sec:conclusion}, we conclude and discuss.


\section{Cosmological scene}  \label{Sec:cosmology}

The logarithm polynomial parameterization was firstly proposed to quantify the deviation between the concordance $\Lambda$CDM cosmology and the high-z QSO Hubble diagram in~\citet{Risaliti:2018reu}. It defines the luminosity distance with a polynomial function of $\log_{10}( 1+z ) $
\begin{equation}
 d_{L}(z)=  \frac{c\ln(10)}{H_0}\sum_{i=1}^{n}a_i\log^{i}_{10}(1+z), \label{eq:logarithm-polynomial}
\end{equation}
where $a_i$ are free parameters, except $a_1=1$. $\rm c$ is the speed of light. $H_0$ is the Hubble constant.

Based on the Taylor expansion in $\log_{10}(1+z)$, this parameterization provides a model-independent exploration of the late-time cosmological expansion history and approximates many cosmological models accurately at low redshift ($z\lesssim1$). However, its approximation precision worsens considerably when $z$ exceeds 1, as shown in Table~\ref{tab:fitting_precision} ( here we take its 4th-order expansion as an example ). For the redshift range $[0, 8]$, the maximum relative errors in luminosity distance are more than $2\%$ for different $\Lambda$CDM models,  which may cause biases in data fitting.  Introducing higher orders in $\log_{10}(1+z)$ could certainly improve the fitting precision but also weaken the constraint power of data and complicate the procedure of comparing with the standard $\Lambda$CDM model.

\begin{table*}
\centering 
\caption{ Comparisons of maximum relative errors in luminosity distance ($d_L$) between 4th-order logarithmic polynomial expansion and PAge approximation. }
\label{tab:fitting_precision}
\begin{tabular}{|c|c|c|c|c|c|c|c|}
\hline\hline
 fiducial cosmology&- &4th-order logarithmic polynomial expansion& PAge approximation\\
\hline
$\Lambda$CDM &redshift ranges& max$|\frac{\Delta d_L}{d_L}|$& max$|\frac{\Delta d_L}{d_L}|$\\ \hline
\\
 \multirow{7}{*}{$\Omega_m=0.3$}&[0,1]& $7.33\times{10}^{-5}$ &$3.82\times{10}^{-4}$ \\ \\
  &[0,2]&$1.74\times{10}^{-3}$ &$1.07\times{10}^{-3}$  \\ \\
  &[0,4]&$1.04\times{10}^{-2}$ &$2.00\times{10}^{-3}$  \\ \\
   &[0,8]&$3.72\times{10}^{-2}$ &$2.85\times{10}^{-3}$  \\ \\
  \hline
  \\
  {$\Omega_m=0.5$}&[0,8]& $2.61\times{10}^{-2}$&$9.86\times{10}^{-4}$ \\ \\
  \hline
  \\
  {$\Omega_m=0.7$}&[0,8]& $2.04\times{10}^{-2}$&$2.68\times{10}^{-4}$  \\ \\
  
 \hline\hline
\end{tabular}
\end{table*}

Compared to the logarithm polynomial approximation, the recently proposed PAge approximation displays many prominent advantages in blindly modelling the late cosmological expansion history~\citep{Huang:2020mub,Luo:2020ufj,Huang:2020evj,Huang:2021aku,Huang:2021tvo,Cai:2021weh,Cai:2022dkh, Huang:2022txw}.  Faithfully obeying the asymptotic matter-dominated assumption $\frac{1}{1+z}\propto t^{\frac{2}{3}}$ at high redshift $z\gg1$ ( the radiation component is not taken into consideration), PAge models the Hubble expansion rate as a function of cosmological time t,
\begin{equation}
\frac{H}{H_0} = 1+\frac{2}{3} (1-\eta \frac{ H_{0}t}{p_{\rm age}} )(\frac{1}{ H_{0}t}-\frac{1}{ p_{\rm age}}), 
\label{eq:Hubblerate}
\end{equation}
where the dimensionless parameter $p_{\rm age} \equiv  H_{0}t_{0} $ measures the cosmic age $t_0$ (both t and $t_0$ are in unit of $H^{-1}_{0}$), and the dimensionless parameter $\eta$ characterizes the deviation from Einstein de-sitter universe (flat CDM model). We set a bound condition $\eta<1$ to guarantee the fundamental physical features,  e.g. $\frac{\mathrm{d} d_{L}}{\mathrm{d} z}> 0$ and $\frac{\mathrm{d} H}{\mathrm{d} z}> 0$~\citep{Huang:2020mub}.

\begin{figure}
\centering
\includegraphics[width=0.48\textwidth]{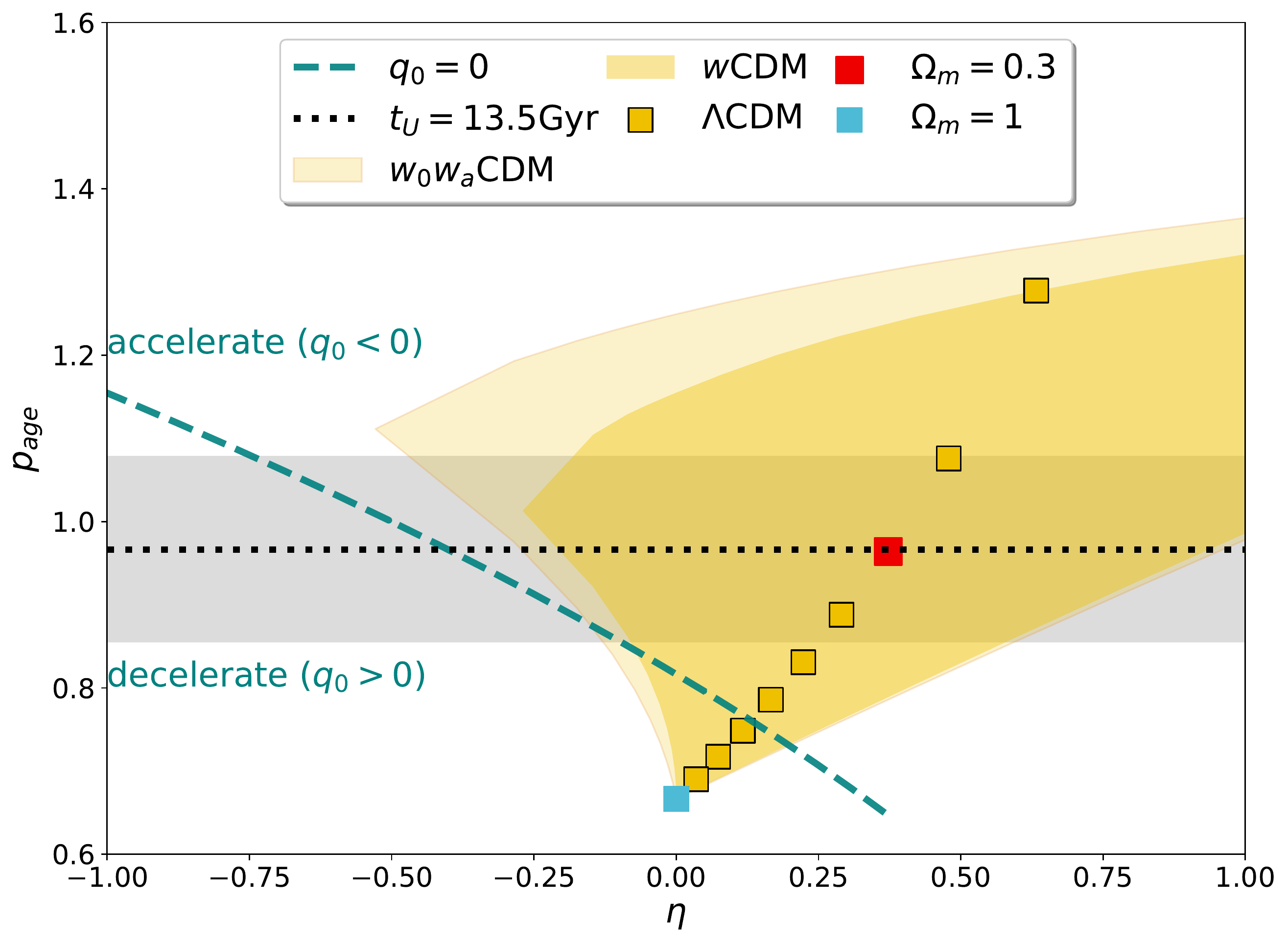}
\caption{Mapping the $\Lambda$CDM, $w$CDM and $w_0$-$w_a$CDM models into $(p_{\rm age}, \eta)$ plane. We map the cosmological models into $(p_{\rm age}, \eta)$ plane by matching the deceleration parameter $q_0$ at $z=0$~\citep{Luo:2020ufj}. The $\Lambda$CDM models are mapped into a sequence of discrete points ($\Omega_m\in[0.1,1]$ with step length 0.1). The $w$CDM models ($\Omega_m\in[0.1, 1]$, $w\in[-1.5, -0.5]$) and $w_0$-$w_a$CDM models ($\Omega_m\in[0.1, 1]$, $w_0\in[-1.5, -0.5]$, $w_a\in[0.5, 1.5]$) are mapped into yellow areas. The green dashed line divides the  $(p_{\rm age}, \eta)$ plane into cosmic deceleration and acceleration regions. The black dotted line and grey region represent the cosmic age and its $3\sigma$ bound inferred in \citet{Valcin:2020vav}. The red point and blue point are the standard $\Lambda$CDM and CDM models.} 
\label{fig:Mapping_models}
\end{figure}

 Doing cosmological tests with PAge approximation has some distinct advantages, as follows:
 \begin{enumerate}
   
     \item [1)] The cosmic age $t_0$ absorbed in $p_{\rm age}$ parameter is easily applied to do astronomical tests~\citep{Luo:2020ufj}. More specifically,  \citet{Valcin:2020vav} presented an independent inference of $t_0=13.5^{+0.16}_{-0.14}(\rm stat.)\pm 0.5(\rm sys.)$ from the full colour-magnitude diagram of the globular cluster. If a cosmological model predicts a significantly different $t_0$ compared to the above estimation, it can be ruled out safely. For instance, the flat CDM model corresponding to $p_{\rm age}=\frac{2}{3}, \eta=0$ in PAge approximation fails to accommodate this cosmic age inference, which is clearly shown in Figure~\ref{fig:Mapping_models}.  \\
     
     \item [2)] The cosmic deceleration and acceleration are easy to distinguish in PAge. According to $\eta=1-\frac{3}{2}p^{2}_{\rm age}(1+q_0)$~\citep{Luo:2020ufj},  the PAge universe is divided into decelerating and accelerating regions in Figure~\ref{fig:Mapping_models}. \\
     
     \item [3)] As an almost model-independent framework, PAge is able to precisely approximate a broad class of physical models by matching the deceleration parameter $q_0$ or by doing a least-square fitting of cosmological observables~\citep[see][]{Huang:2020mub, Luo:2020ufj, Huang:2020evj, Huang:2021aku}. The maximum relative errors of luminosity distance ($d_L$) are controlled below $0.3\%$, and it is well held for both low redshift and high redshift, as indicated in Table~\ref{tab:fitting_precision}.   \\
     
     \item [4)] Utilizing PAge approximation to do Bayesian analysis is economical, effective, and concise. Generally, many typical physical models can be approximately mapped into the $(p_{\rm age}, \eta)$ plane, and some of them are superimposed onto one point (we visualize this unique feature in Figure~\ref{fig:Mapping_models}). Performing data analysis with PAge provides the Bayesian evidence for all the models which are included in the marginalized contour of $p_{\rm age}$ and $\eta$ parameters. This practice avoids the cumbersome and complex process of computing Bayesian evidence for all the models. 
 \end{enumerate}

 Almost having the same advantages as the logarithm polynomial approximation~\citep{Yang:2019vgk}, PAge merely has two nuisance parameters ($p_{\rm age},\eta$) and simultaneously has reliable fitting precision at both low-z and high-z. Since PAge displays many superiorities and can accurately describe the expansion history of the high-z universe, we utilize it to do an independent analysis of QSO cosmology.

\section{Data And Methodology } \label{Sec:data-method}

A high-quality QSO sample is recently compiled in ~\citet{Lusso:2020pdb}. This new sample includes 2421 optically selected QSOs with spectroscopic redshift (span the redshift interval $0.009\le z\le 7.5413$) and X-ray observations. Systematic effects and low-quality measurements are largely removed by applying a couple of preliminary filters. For example, $30\%$ X-ray measurements are excluded by the conditions: $\Delta F_s/F_s<1$ and $\Delta F_H/F_H<1$. More detailed filter procedures are discussed in ~\citet{Lusso:2020pdb}.  After an optimal selection of clean sources, this high-quality QSO sample is suitable for  investigating the non-linear relation between the ultraviolet (at 2500 \AA, $L_{\rm UV}$) and X-ray (at 2$\rm keV$, $L_{X}$) luminosity of QSO:
\begin{equation}
\log_{10}{L_{\rm X}} =\gamma \log_{10}{L_{\rm UV}}+\beta , 
\label{eq:QSOluminositycorrelation}
\end{equation}
where $\gamma, \beta$ are free parameters. $L_{X}$ and $L_{\rm UV}$ are the rest-frame monochromatic luminosities which follow the standard luminosity-flux relation $L=4\pi d^2_{L}F$. Further expressing Eq.~\ref{eq:QSOluminositycorrelation} with flux, one obtains
\begin{equation}
\log_{10}{F_X}=\gamma\log_{10}{F_{\rm UV}}+(\gamma-1)\log_{10}{(4\pi d^2_L)}+\beta, 
\label{eq:QSOfluxcorrelation}
\end{equation}
both $F_X$ and $F_{\rm UV}$ are the flux densities in the unit of $\rm erg/s/cm^2$. We quantify the uncertainties of $\gamma, \beta$ and the variability of cosmologies with the joint likelihood function~\citep{DAgostini:2005mth}:
\begin{equation}
\ln \mathcal{L} \propto -\frac{1}{2} \sum_{i=1}^{N}  [\frac{(\log_{10}{F^{\rm obs}_{X,i}}-\log_{10}{F^{\rm th}_{X,i}})^2}{
\sigma^2_{\rm total}}+\ln{(2 \pi\sigma^2_{\rm total}})], 
\label{eq:Likelihood}
\end{equation}
the total uncertainties $\sigma^2_{\rm total}=\sigma^2_{F^{\rm obs}_{X,i}}+\gamma^2\sigma^2_{F^{\rm obs}_{\rm UV,i}}+\delta^2$, where $\delta $ is a scatter parameter representing uncounted extra variability.

It is worth noting that the QSOs can not be used to do cosmological tests directly because they do not provide absolute distance values~\citep{Bargiacchi:2021hdp}. A cross-calibration procedure is needed to match the distance values between QSOs and SNe in the common redshift range. The detailed calibration procedure is to multiply the luminosity distance by a calibration parameter $k$, i.e.  $d^{\bf calibration}_{L}(z)=k d^{\bf model}_L(z)$, and the $k$ parameter requires a simultaneous fitting of QSOs and SNe. In the cross-calibration procedure, $k$ degenerates with $H_0$. To avoid parameter degeneracy, we fix $H_0=70 \rm km/s/Mpc$ in the following analyses. We calibrate the QSO distances with the Pantheon SNe sample~\citep{Pan-STARRS1:2017jku} and use the combination data set of SNe+QSOs to build the QSO Hubble diagram.

As indicated in Table 2 of~\citet{Bargiacchi:2021hdp}, the spatial curvature has a great impact on the cosmological constraints when using QSO data. The joint analysis of SNe+QSO in non-flat $\Lambda$CDM background prefers a closed universe with spatial curvature  $\Omega_k\simeq-0.6$, which is inconsistent with the Planck+CMB result $\Omega_k=-0.044^{+0.018}_{-0.015}$~\citep{Planck:2018vyg}. To not miss some important information, we consider both flat and non-flat cosmological cases in the following analyses.

\section{ Results}\label{Sec:results}

\begin{table*}
\centering 
\caption{  Marginalized $1\sigma$ constraints on parameters with SNe+QSOs samples in PAge backgrounds. }
\label{tab:constraints}
\begin{tabular}{|c|c|c|c|c|c|c|c|}
\hline\hline
model & $\Omega_k$& $p_{\rm age}$ & $\eta$ &$ \gamma$ & $\beta$ & $\delta$ \\ \hline
\\
flat PAge &-&$0.820^{+0.017}_{-0.02}$&$0.896^{+0.089}_{-0.038}$ & $0.642\pm0.0083$&$6.96^{+0.33}_{-0.29}$&$0.228\pm0.0034$\\
\\
non-flat PAge&$-0.946^{+0.017}_{-0.051}$ & $1.16\pm0.051$ & $0.540^{+0.17}_{-0.14}$ &$0.619\pm0.009$ &$7.68\pm0.33$&$0.226\pm0.0036$\\

 \\ \\ \hline\hline
\end{tabular}
\end{table*}

\begin{figure}
\centering
\includegraphics[width=0.48\textwidth]{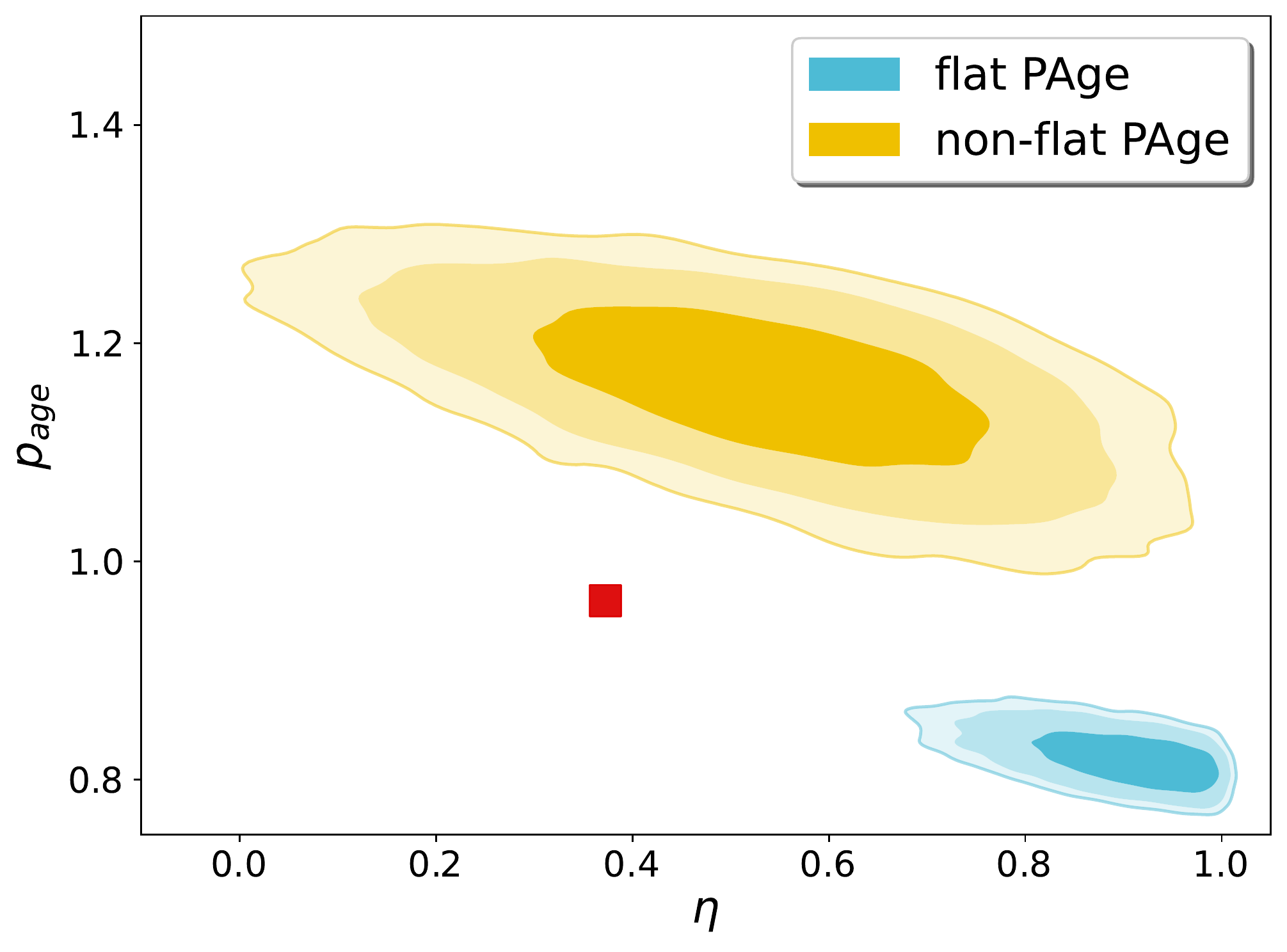}
\caption{Marginalized 1$\sigma$, 2$\sigma$ and 3$\sigma$ constraints on PAge parameters with  SNe+QSOs samples. Both flat and non-flat PAge universes are taken into account. The red point represents the standard $\Lambda$CDM which corresponds to the point ($p_{\rm age}=0.964$, $\eta=0.373$) in PAge panel.}
\label{Fig:contour}
\end{figure}

In Table~\ref{tab:constraints}, we list the marginalized 1$\sigma$ constraints on parameters with SNe+QSOs data in PAge backgrounds. Both the flat and non-flat cases are taken into consideration. For better comparison with the standard $\Lambda$CDM model, we visualize the marginalized 1$\sigma$, 2$\sigma$ and 3$\sigma$ constraints on PAge parameters in Figure~\ref{Fig:contour}.  We find both the marginalized contours on $p_{\rm age}$ and $\eta$ in flat and non-flat PAge cases significantly deviate from the standard $\Lambda$CDM model ( red point in Figure~\ref{Fig:contour} ) at $>3\sigma$ confidence level. The marginalized contours of the flat and non-flat PAge cases also show a $>3\sigma$ discrepancy, which indicates the inferences of PAge parameters are much affected by the addition of spatial curvature $\Omega_k$ freedom. Indeed, the spatial curvature $\Omega_k$ is found to be $-0.946^{+0.017}_{-0.051}$, which strongly supports a closed universe. And the exotic $\Omega_k$ inference is actually inconsistent with other measurements~\citep{Pan-STARRS1:2017jku, eBOSS:2020yzd, Jimenez:2001gg, DES:2021wwk, Planck:2018vyg}.

 The SNe+QSOs constraints on PAge parameters in both flat and non-flat universes seem to suggest new physics beyond the $\Lambda$CDM. However, whether the QSOs can serve as a reliable cosmological tool still requires cautious research. As mentioned in~\citet{Lusso:2020pdb}, using the non-linear QSO luminosity correlation to build the QSO Hubble diagram may still have shortcomings. For example, the systematics in the QSO samples selection, the process used to fit the QSO Hubble diagram and the redshift evolution effect of QSO luminosity correlation may cause biases and lead to an unreliable QSO Hubble diagram. Possible systematics have been carefully checked in~\citet{Lusso:2020pdb} and we focus on the redshift evolution effect of the QSO luminosity correlation in this present work.

\begin{table*}
\centering 
\caption{ Marginalized $1\sigma$ constraints on parameters with SNe+low-z QSOs sample and SNe+high-z QSOs sample respectively.} 
\label{tab:redshift_evolution}
\begin{tabular}{|c|c|c|c|c|c|c|c|c|}
\hline\hline
model &sample&  $\Omega_k$ & $p_{\rm age}$ & $\eta$ &$ \gamma$ & $\beta$ & $\delta$ \\ \hline
\\
  \multirow{3}{*}{flat PAge}&SNe+low-z QSOs& -&$0.940^{+0.033}_{-0.062}$ & $0.480^{+0.21}_{-0.13}$ & $0.637\pm0.013$ & $7.18\pm0.41$ & $0.237\pm0.0047$ \\ \\
  &SNe+high-z QSOs& -&$0.875^{+0.025}_{-0.032}$ & $0.727^{+0.12}_{-0.096}$ & $0.585\pm0.014$ & $8.80^{+0.52}_{-0.43}$ & $0.207\pm0.0054$ \\ \\ \hline
  \\
  \multirow{3}{*}{non-flat PAge}&SNe+low-z QSOs&  $-0.512^{+0.096}_{-0.35}$ &$1.21^{+0.28}_{-0.10}$ & $0.05^{+0.48}_{-0.35}$ & $0.635\pm0.013$ & $7.21\pm0.43$ & $0.237\pm0.0047$ \\ \\
  &SNe+high-z QSOs&  $-0.795\pm0.078$  &$1.24^{+0.057}_{-0.071}$ & $0.220^{+0.26}_{-0.19}$ & $0.561^{+0.0088}_{-0.013}$ & $9.49^{+0.49}_{-0.15}$ & $0.203\pm0.0053$ \\ \\ \hline\hline
\end{tabular}
\end{table*}

Different from the narrow redshift bins split in~\citet{Risaliti:2018reu,Lusso:2020pdb}, we divide QSO samples into the low-z ($z\le1.5$) bin and high-z bin ($z>1.5$) to test the redshift evolution, because the strong deviation from $\Lambda$CDM model roughly emerges at $z>1.5$. We analyze the SNe+low-z QSOs and SNe+high-z QSOs data sets in flat and non-flat PAge backgrounds, respectively. The marginalized $1\sigma$ constraints on the PAge parameters, QSO correlation coefficients and intrinsic dispersion are presented in Table~\ref{tab:redshift_evolution}.

\begin{figure*}
\centering
\subfigure{
	\includegraphics[width=0.48\textwidth]{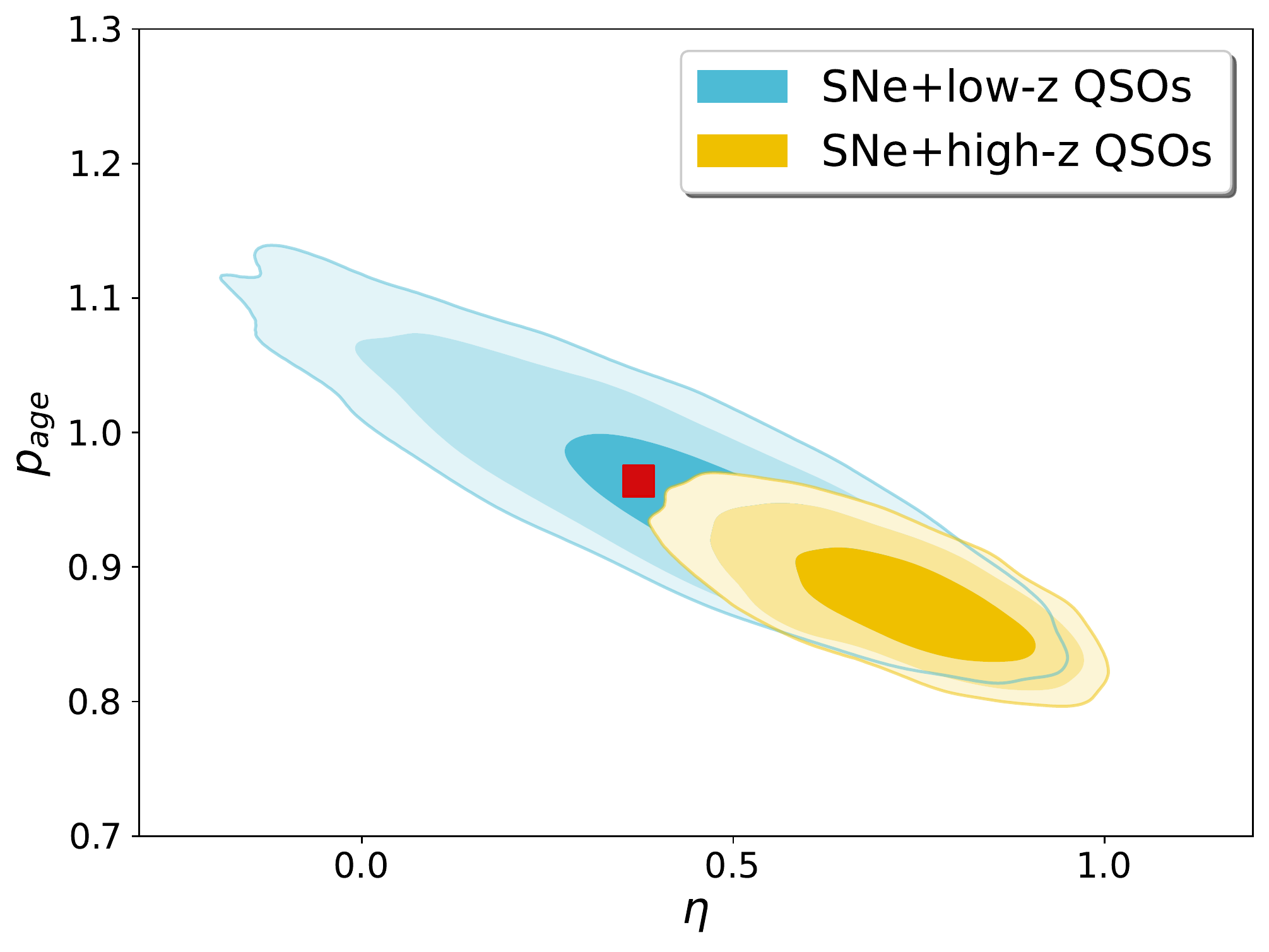}}
\subfigure{
	\includegraphics[width=0.48\textwidth]{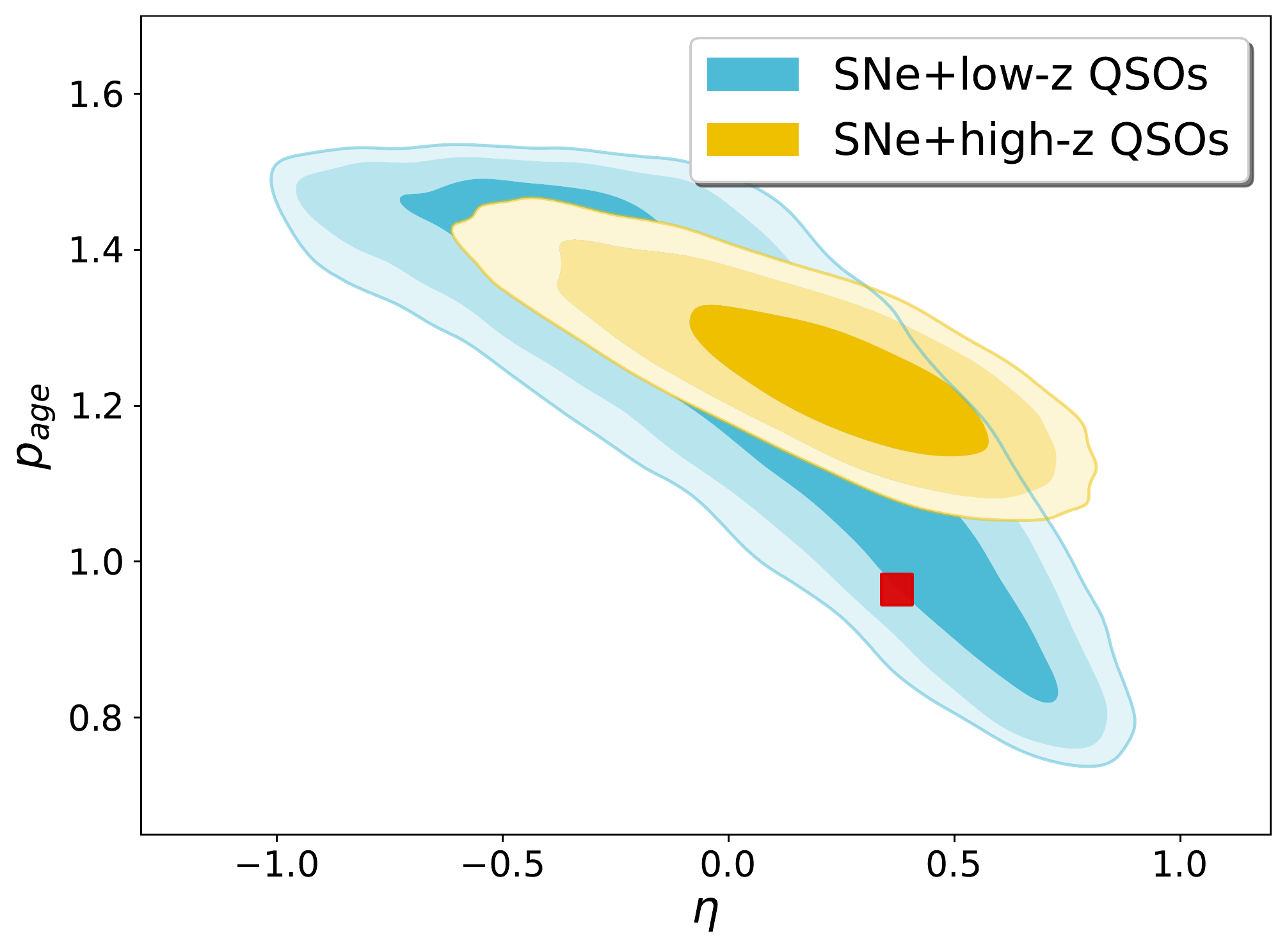}}
\caption{Marginalized $1\sigma$, $2\sigma$ and $3\sigma$ constraints on PAge parameters with SNe+low-z QSOs and SNe+high-z QSOs samples in flat and non-flat PAge universes, respectively. The red point represents the standard $\Lambda$CDM model.} 
\label{fig:comparison_D}
\end{figure*}

Either in flat or in non-flat PAge cases, significant evolutionary trends emerge for the $\gamma, \delta$ parameters. For the $\gamma$ parameter, we find $\sim 2.7\sigma$ and $\sim 4\sigma$ discrepancies between the two data sets in flat and non-flat PAge universes respectively. For the $\delta$ parameter, the discrepancies are found to be $>4\sigma$ in different PAge backgrounds. These results indicate that the QSO luminosity correlation evolves with redshift and suffers from non-universal dispersion. In addition, we find the PAge parameters inferred from the SNe+low-z QSOs and SNe+high-z QSOs data sets, as shown in Figure~\ref{fig:comparison_D}, coincide well. More importantly, the marginalized $3\sigma$ contour of  SNe+high-z QSOs deviates from the $\Lambda$CDM significantly while the SNe+low-z QSOs case roughly accommodates it, which implies that the PAge parameter inferences are probably biased by the redshift evolution and non-universal dispersion.

\begin{table*}
\centering 
\caption{ Marginalized $1\sigma$ constraints on parameters with Reichart method.} 
\label{tab:Reichart method}
\begin{tabular}{|c|c|c|c|c|c|c|c|c|}
\hline\hline
model &sample&  $\Omega_k$ & $p_{\rm age}$ & $\eta$ &$ \gamma$ & $\beta$ & $\delta$ \\ \hline
\\
  \multirow{3}{*}{flat PAge}&SNe+low-z QSOs& -&$1.05^{+0.051}_{-0.10}$ & $0.140^{+0.36}_{-0.18}$ & $0.750^{+0.013}_{-0.011}$ & $3.72^{+0.32}_{-0.46}$ & $0.244\pm0.0050$ \\ \\
  &SNe+high-z QSOs& -&$0.919^{+0.033}_{-0.044}$ & $0.570^{+0.16}_{-0.11}$ & $0.681\pm0.016$ & $5.84\pm0.51$ & $0.213\pm0.0059$ \\ \\ \hline
  \\
  \multirow{3}{*}{non-flat PAge}&SNe+low-z QSOs&  $-0.240^{+0.12}_{-0.44}$ &$1.20^{+0.29}_{-0.10}$ & $-0.096^{+0.52}_{-0.39}$ & $0.748^{+0.014}_{-0.012}$ & $3.76^{+0.35}_{-0.44}$ & $0.244\pm0.0050$ \\ \\
  &SNe+high-z QSOs&  $-0.638^{+0.081}_{-0.10}$  &$1.26\pm0.11$ & $-0.01^{+0.38}_{-0.26}$ & $0.660\pm0.017$ & $6.45\pm0.55$ & $0.210\pm0.0058$ \\ \\ \hline\hline
\end{tabular}
\end{table*}

In the Bayesian framework, performing a linear fit between two data sets with errors on both axes and with an extra variance is quite subtle.  Different analysis methods may yield inconsistent results, as indicated in~\citet{Guidorzi:2006wb}. In our above analyses, we use the likelihood function~(\ref{eq:Likelihood}) derived by~\citet{DAgostini:2005mth} to estimate the parameters and find remarkable discrepancies for $\gamma, \delta$ parameters.  To demonstrate the discrepancies are not dominated by the statistical analysis method, we further perform Bayesian analyses with the Reichart method~\citep{2001ApJ...553..235R, 2001ApJ...552...57R} and present the parameter inference results in Table~\ref{tab:Reichart method}.

According to Table~\ref{tab:Reichart method}, we find the $\gamma, \delta$ parameters derived from the Reichart method also show prominent discrepancies. The $\gamma$ parameters calibrated by SNe+low-z QSOs samples are in $\sim 3.4\sigma$ and $\sim 4.1\sigma$ tension with that calibrated by SNe+high-z QSOs samples in flat and non-flat PAge respectively. The $\delta$ parameters  show $\sim4.0\sigma$ and $\sim 4.4\sigma$ discrepancies between different data sets and backgrounds. This suggests that the redshift-evolution effect and non-universal dispersion of QSO luminosity correlation are independent of the statistical analysis method.


\section{Conclusions and Discussion}  \label{Sec:conclusion}

In this research, we provide an independent search for the origins of the $\sim4 \sigma$ deviation between the standard $\Lambda$CDM model and the constructed Hubble diagram of  SNe+QSOs~\citep{Risaliti:2018reu, Lusso:2019akb, Lusso:2020pdb}. We adopt a nearly model-independent parameterization (PAge approximation) to visualize the standard $\Lambda$CDM model and marginalized $3\sigma$ constraints of  SNe+QSOs data. To a certain degree, we have avoided the model dependence and the fitting errors of the assumed background cosmology~\citep{Yang:2019vgk}. According to the results shown in Figure~\ref{Fig:contour}, we confirm that the marginalized $3\sigma$ constraints of SNe+QSOs on PAge parameters are in remarkable tension with the standard $\Lambda$CDM model in both flat and non-flat universes. This result agrees with~\citet{Risaliti:2018reu, Lusso:2019akb, Lusso:2020pdb}.

We proceed to investigate the tension from the perspective of redshift evolution. By splitting QSOs into low-z and high-z samples, we find that there indeed exist remarkable discrepancies for the slope $\gamma$ parameter and intrinsic dispersion $\delta$ between low-z and high-z QSOs calibrated by SNe. And the remarkable discrepancies for $\gamma$ and $\beta$ parameters persist in the parameter inferences derived from the Reichart method. These results reveal that the QSO luminosity correlation suffers from the redshift-evolution effect and non-universal intrinsic dispersion.

Building a QSO Hubble diagram with a non-robust QSO luminosity correlation may provide unreliable results. As indicated in Figure~\ref{fig:comparison_D}, with the evolutions of the $\gamma$ and $\delta$ parameters, the marginalized contour of the SNe+low-z QSOs sample is consistent with the $\Lambda$CDM while significant deviation emerges for the SNe+high-z QSOs case. This indicates the PAge parameter constraints can be biased by the evolutions of $\gamma$ and $\delta$ parameters. Therefore, the significant deviation found in~\citep{Risaliti:2018reu, Lusso:2019akb, Lusso:2020pdb} may mainly originate from the redshift-evolution effect and the non-universal intrinsic dispersion of the QSO luminosity correlation instead of new physics beyond the $\Lambda$CDM cosmology.

\section*{Acknowledgements}

We gratefully thank the authors in~\citet{Lusso:2020pdb} for sharing the data used in this work. We thank Zhiqi Huang for the helpful discussions. This work is supported by the National Natural Science Foundation of China (NSFC) under Grant No. 12073088, National SKA Program of China No. 2020SKA0110402, National key R\&D  Program of China (Grant No. 2020YFC2201600), and Guangdong Major Project of Basic and Applied Basic Research (Grant No. 2019B030302001).


\section{Data Availability}
The quasar data underlying this article are available in~\citet{Lusso:2020pdb} and in its online supplementary material. The supernova data underlying this article are publicly available in~\citet{Pan-STARRS1:2017jku}. 



\bibliographystyle{mnras}
\bibliography{ref} 





\label{lastpage}
\end{document}